\documentclass[a4paper]{jpconf}
\usepackage{graphicx}
\usepackage{amsmath}
\usepackage{amsfonts}
\usepackage{braket}
\begin{document}
\title{Fermion mixing in curved spacetime}

\author{Antonio Capolupo, Gaetano Lambiase and Aniello Quaranta}

\address{Dipartimento di Fisica “E.R. Caianiello” Universita di Salerno,
and INFN – Gruppo Collegato di Salerno, Via Giovanni Paolo II, 132, 84084 Fisciano (SA), Italy}

\ead{capolupo@sa.infn.it, lambiase@sa.infn.it, anquaranta@unisa.it}

\begin{abstract}
 We develop the quantum field theory of fermion
mixing in curved spacetime and discuss the role
of unitarily inequivalent representations in the
particle interpretation of the theory. We derive
general oscillation formulae and apply them to
specific spcetimes of interest, such as spatially
flat FRW metrics and the Schwarzschild spacetime. We exhibit the main deviations from the
usual quantum mechanical approach.
\end{abstract}

\section{Introduction}

It is today accepted that neutrinos have a mass and oscillate among three flavors \cite{neut1,neut2,neut3,neut4,neut5,neut6,neut7,neut8,neut9,neut10,neut11,neut12,neut13,neut14}. They are the only known elementary particles to experience field mixing. These peculiarities place neutrinos beyond the standard model of particles. Many of the issues related to neutrino physics, including the origin of their mass \cite{mass1}, their fundamental nature \cite{nat1,nat2,nat3,nat4,nat5} and the overall number of flavors \cite{flav1} are still open. It is nevertheless clear that neutrinos play a fundamental role in the universe at all the scales. As they are abundantly produced in nuclear reactions, they carry important informations on astrophysical sources and they represent a valuable resource in multi-messenger astronomy \cite{MMA}. Relic neutrinos, as probed by experiments like PTOLEMY \cite{PTOLEMY}, may be used to test cosmological theories. Several theories envisage a crucial role of neutrinos during the first phases of the universe in producing the original baryon asymmetry \cite{Lepto}. Furthermore neutrinos may be connected to the dark sector of the universe, contributing to dark matter \cite{DM1,DM2,DM3,DM4,DM5} along with hypothetical particles such as axions \cite{DMA1,DMA2,DMA3}, and even to dark energy \cite{DE1,DE2,DE3,DE4}, for which they can also function as a probe of the underlying model \cite{DEP1,DEP2,DEP3}. Due to these reasons, a deep understanding of neutrino physics in gravitational backgrounds is required. The topic has been analyzed primarily in quantum mechanical approaches \cite{QMO1,QMO2,QMO3} both in vacuum and in matter \cite{QMO3,QMM1,QMM2,QMM3}. Here we wish to go beyond the quantum mechanical treatment and present a quantum field theoretical approach to fermion mixing in curved space. We limit ourselves to two flavors, but the formalism can be easily generalized to more families. The theory merges the features of mixing with the field quantization on curved space, bringing along several interesting aspects, such as a non-trivial structure of the vacuum. We proceed as follows: in Section 2 we introduce the flavor fields and pursue their canonical quantization; in Section 3 we derive the transition probabilities for a generic spacetime; in Section 4 we apply the formalism to some specific metrics and compute the corresponding oscillation formulae, showing their departure from their quantum mechanical counterparts; Section 5 is devoted to the conlcusions.      

\section{Flavor fields in curved spacetime}

We start with the Dirac equations for the fields with definite masses
\begin{equation}\label{DiracEquations}
 \left(i \gamma^{\mu}(x) D_{\mu} - m_j \right) \nu_j = 0 \ ,
\end{equation}
where $j=1,2$, $D_{\mu} = \partial_{\mu} - \frac{i}{4} \omega^{AB}_{\mu} \sigma_{AB}$ is the spinorial covariant derivative and $\gamma^{\mu} (x) = e^{\mu}_A (x) \gamma^A$ are the curved space gamma matrices. It is understood that a tetrad $e^{\mu}_A (x)$, with Lorentz index $A = 0,1,2,3$ , is picked on the manifold, which is therefore assumed parallelizable. We will also require that the underlying manifold be globally hyperbolic, thus admitting a foliation by Cauchy surfaces $\Sigma$. The spin connection is accordingly given by $\omega^{AB}_{\mu} = e^A_{\nu}\Gamma^{\nu}_{\rho \mu} e^{\rho B} + e^{A}_{\nu}\partial_{\mu} e^{\nu B}$, with $\Gamma^{\nu}_{\rho \mu}$ the Christoffel symbols, and coupling to the (spacetime constant) coefficients $\sigma_{AB} = \frac{i}{2} \left[ \gamma_A, \gamma_B \right]$. The Lagrangian density corresponding to Equations \eqref{DiracEquations} is
\begin{equation}\label{DiracLagrangian}
 \mathcal{L} = \sqrt{-g} \sum_{j=1,2} \bar{\nu}_j \left(i \gamma^{\mu}(x) D_{\mu} - m_j  \right) \nu_j \ .
\end{equation}
Notice that the above Lagrangian is evidently invariant under global $U(1)$ gauge transformations $\nu_j \rightarrow e^{i \beta} \nu_j$. 
For a given manifold we can solve the Equations \eqref{DiracEquations} for ``positive energy'' $\phi_{k,s,j} (x)$ and ``negative energy'' $\psi_{k,s,j} (x)$ modes. Here $k,s$ are generalized momentum and spin indices, while $j=1,2$ distinguishes the two massive fields. A crucial point is that on a generic spacetime there is, in general, no global or natural definition of ``positive energy modes''. One has to pick a timelike vector field $\frac{\partial}{\partial T}$ and define the energy with respect to it. While in Minkowski space this field is essentially unique, the same does not hold in a generic spacetime. Such an ambiguity is the source of the well-known particle creation phenomena in curved space (Parker effect \cite{Parker}, Hawking radiation \cite{Hawking}). Given a set of solutions $\phi_{k,s,j}, \psi_{k,s,j}$, which is complete with respect to the inner product
\begin{equation}\label{InnerProduct}
 (f, g)_{\tau} = \int_{\Sigma (\tau)} \sqrt{-g} \ d \Sigma_{\mu} (\tau) \bar{f} \gamma^{\mu} (x) g 
\end{equation}
we can expand the fields $\nu_j$ as usual. Here $\tau \in \mathbb{R}$ labels the foliation by Cauchy surfaces, $g = \det{g_{\mu \nu}}$ and the integral is performed over the surface $\Sigma (\tau)$ with (3-)volume element $d \Sigma_{\mu} (\tau)$. We anticipate that the inner product between solutions of the same Dirac equation (that is, with the same mass) does not depend on $\tau$, as a consequence of the $U(1)$ symmetry of the Lagrangian \eqref{DiracLagrangian}. On the other hand the product between solutions of distinct Dirac equations generally depends on $\tau$. The fields are then expanded as 
\begin{equation}\label{FreeFieldExpansion}
 \nu_j (x) = \sum_s \int d^3 k \left( a_{k,s;j} \phi_{k,s,j} (x) + b^{\dagger}_{k,s;j} \psi_{k,s,j}(x)\right)
\end{equation}
with the (spacetime constant) coefficients $a,b$ satisfying the canonical anticommutation relations $\left \lbrace a_{k,s;j}, a^{\dagger}_{q,r;l} \right \rbrace = \delta_{jl} \delta_{sr} \delta_{kq} = \left \lbrace b_{k,s;j}, b^{\dagger}_{q,r;l} \right \rbrace$. The next step is to introduce the flavor fields by means of an $SU(2)$ rotation through the mixing angle $\theta$
\begin{eqnarray}
\nonumber \nu_e (x) &=& \cos \theta \ \nu_1 (x) + \sin \theta \ \nu_2 (x) \\
 \nu_{\mu} (x) &=& -\sin \theta \ \nu_1 (x) + \cos \theta \ \nu_2 (x) \ ,
\end{eqnarray}
or, which is equivalent, by means of the \emph{generator of mixing transformations} $\mathcal{I}_{\theta} (\tau)$
\begin{equation}\label{Generator1}
 \nu_e (x) = \mathcal{I}_{\theta}^{-1}(\tau) \nu_1 (x) \mathcal{I}_{\theta} (\tau) \ ; \ \ \ \ \nu_{\mu} (x) = \mathcal{I}_{\theta}^{-1}(\tau) \nu_2 (x) \mathcal{I}_{\theta} (\tau) \ .
\end{equation}
Inspection of the above equations reveals that the generator can be written as
\begin{equation}\label{Generator2}
 \mathcal{I}_{\theta} (\tau) = \exp \left \lbrace \theta \left[\left(\nu_1, \nu_2\right)_{\tau} - \left(\nu_2, \nu_1\right)_{\tau}\right] \right \rbrace \ .
\end{equation}
The application of the generator on the mass fields defines the (surface-wise) expansion of the flavor fields, as 
\begin{eqnarray}\label{FlavorFieldExpansion}
\nonumber \nu_e (x) &=& \sum_s \int d^3 k \left(a_{k,s;e} (\tau) \phi_{k,s;1} (x)+ b^{\dagger}_{k,s;e}(\tau) \psi_{k,s;1}  (x) \right) \\ 
\nu_{\mu} (x) &=& \sum_s \int d^3 k \left(a_{k,s;\mu} (\tau) \phi_{k,s;2} (x)+ b^{\dagger}_{k,s;\mu}(\tau) \psi_{k,s;2}  (x) \right) \ ,
\end{eqnarray}
where the flavor operators are by definition 
\begin{equation}\label{FlavorOperators}
 a_{k,s;e} (\tau) = \mathcal{I}^{-1}_{\theta} (\tau) a_{k,s;1} \mathcal{I}_{\theta} (\tau) = \cos \theta a_{k,s;1} +  \sin \theta \sum_{r} \int d^3 q \left(\Lambda^*_{q,r;k,s}(\tau) a_{q,r;2} + \Xi_{q,r;k,s} (\tau) b^{\dagger}_{q,r;2} \right) 
\end{equation}
and similar for $a_{\mu}, b_{e}, b_{\mu}$. The \emph{Bogoliubov coefficients of the mixing transformations} are given by the inner products
\begin{equation}\label{BogoliubovCoefficients1}
 \Lambda_{q,r;k,s} (\tau) = \left(\phi_{q,r,2}, \phi_{k,s,1} \right)_{\tau} \ ; \ \ \ \ \Xi_{q,r;k,s} (\tau) = \left(\psi_{k,s,1}, \phi_{q,r,2} \right)_{\tau} \ .
\end{equation}
That these coefficients define a fermionic Bogoliubov (linear canonical) transformation is ensured by the property 
\begin{equation}
 \sum_{r} \int d^3 q \left( \Lambda^*_{k,s;q,r} (\tau) \Lambda_{k',s', q,r} (\tau) + \Xi^*_{k,s;q,r} (\tau) \Xi_{k',s', q,r} (\tau)\right) = \delta_{kk'}\delta_{ss'} \ .
\end{equation}
The expansions of the free fields \eqref{FreeFieldExpansion} and of the flavor fields \eqref{FlavorFieldExpansion} define two distinct Fock space representations, with the mass vacuum $\ket{0_M}$ and the flavor vacuum $\ket{0_F (\tau)}$ satisfying respectively
\begin{equation}
 a_{k,s;i} \ket{0_M} = 0 = b_{k,s;i} \ket{0_M} \ ; \ \ \ \ a_{k,s;\alpha} (\tau) \ket{0_F(\tau)} = 0 = b_{k,s;\alpha} (\tau) \ket{0_F(\tau)}
\end{equation}
for all $i = 1,2$ and $\alpha = e, \mu$. The appearance of a Bogoliubov transformation in the flavor operators \eqref{FlavorOperators} shows that the mass and the flavor representations are unitarily inequivalent. In particular, the particle content of the two vacua is different and it can be shown \cite{neut10} that the flavor vacuum has the structure of a condensate of particle-antiparticle pairs with definite masses. Notice that both the flavor operators and the flavor vacuum carry an intrinsic $\tau$ dependence, regardless of the set of solutions chosen for the expansion. The one particle states $\ket{\nu_{k,s;\alpha}(\tau)} = a^{\dagger}_{k,s;\alpha} (\tau) \ket{0_F (\tau)}$ are interpreted as the states for a single neutrino of momentum $k$, spin $s$ and flavor $\alpha = e, \mu $. Clearly the $\tau$ dependence of these states is due to the non-conservation of flavor by the Lagrangian \eqref{DiracLagrangian}.

\section{Oscillation Formulae}

In addition to the unitarily inequivalence between mass and flavor representations, it should be stressed that the flavor representations themselves are unitarily inequivalent to each other at distinct times $\tau \neq \tau'$ \cite{neut10,neut13}. The flavor transition amplitudes cannot be introduced by simple scalar products of the form $\bra{\nu_{\alpha}(\tau)} \nu_{\beta} (\tau') \rangle$ because, strictly speaking, they all vanish for $\tau' \neq \tau$, as the states belong to mutually orthogonal Hilbert spaces. Rather we have to identify a suitable operator whose matrix elements give the flavor transition probability. The crucial observation is that due to the $U(1)$ invariance of the Lagrangian \eqref{DiracLagrangian} there exists a conserved (i.e. $\tau$ independent) charge
\begin{eqnarray}\label{Charge}
 \nonumber Q &=& \sum_{j=1,2} Q_j = \sum_{\alpha = e, \mu} Q_{\alpha} (\tau) \ , \\ \mathrm{with} \ \ Q_{\lambda} (\tau) &=& \sum_s \int d^3 k \left( a^{\dagger}_{k,s;\lambda} (\tau) a_{k,s;\lambda}(\tau) - b^{\dagger}_{k,s;\lambda} (\tau) b_{k,s;\lambda}(\tau)\right)
\end{eqnarray}
and $\lambda = j, \alpha$ runs over all possible particle indices $\lambda =1,2,e,\mu$. Equation \eqref{Charge} takes into account the fact that the massive charges $Q_j$ are separately conserved, while the flavor charges $Q_{\alpha}(\tau)$ are not, although their sum is. We \emph{define} the transition probabilities as
\begin{equation}\label{TransitionProbabilities1}
 P_{k,s}^{\alpha \rightarrow \beta} (\tau, \tau_0) = \bra{\nu_{k,s;\alpha}(\tau_0)} Q_{\beta} (\tau) \ket{\nu_{k,s;\alpha}(\tau_0)} - \bra{0_F(\tau_0)}Q_{\beta} (\tau) \ket{0_F (\tau_0)} \ ,
\end{equation}
where $\tau_0$ is a reference time and the subtraction of the second term on the right hand side amounts to the normal ordering with respect to $\ket{0_F (\tau_0)}$. Such a subtraction is required in order that the probabilities be well-defined. In particular, by construction, one has $P^{\alpha \rightarrow \beta } (\tau, \tau_0) \leq 1$, $\sum_{\beta = e, \mu} P^{\alpha \rightarrow \beta} (\tau, \tau_0) = 1$ and $P^{\alpha \rightarrow \beta} (\tau_0 , \tau_0) = \delta^{\alpha \beta}$. It is also easy to show that Equation \eqref{TransitionProbabilities1} defines a proper generalization of the standard oscillation probabilities, to which they reduce when the suitable limits are considered \cite{neut13}. By trivial manipulations we arrive at the explicit expression
\begin{equation}\label{TransitionProbabilities2}
P^{e \rightarrow \mu}_{k,s} (\tau, \tau_0) = \frac{\sin^2 2 \theta}{2} \left[1- \sum_{r} \int d^3 q \ \Re \left(\Lambda^*_{k,s;q,r}(\tau_0) \Lambda_{k,s;q,r}(\tau)  + \Xi^*_{k,s;q,r}(\tau_0) \Xi_{k,s;q,r}(\tau) \right)\right] \ .
\end{equation}
In addition one has $P^{e \rightarrow e}_{k,s} (\tau,\tau_0) = 1 - P^{e \rightarrow \mu}_{k,s} (\tau,\tau_0)=P^{\mu \rightarrow \mu}_{k,s} (\tau,\tau_0)$ and $P^{e \rightarrow \mu}_{k,s} (\tau,\tau_0) = P^{\mu \rightarrow e}_{k,s} (\tau,\tau_0)$. We conclude the section with some remarks about the validity of equations \eqref{TransitionProbabilities1}. It should be noted that the whole construction assumes a given expansion of the mass fields \eqref{FreeFieldExpansion} and thus a specific choice of the set of solutions for the Dirac equation. Clearly one can consider several distinct choices for the mass field representation, and perform the same construction arriving at the transition probabilities of eqs. \eqref{TransitionProbabilities1} and \eqref{TransitionProbabilities2}. In general both the mass vacuum $\ket{0'_M}$ and the corresponding flavor vacua $\ket{0'_F (\tau)}$, as well as the related Fock spaces, shall have different interpretations for different choices of the solutions set. Nonetheless covariance demands that the \emph{local} observables be the same for whatever representation of the mass fields is chosen. This, in turn, implies a specific relation among the flavor representations built out of distinct mass representations \cite{neut13}. On the other hand the transition probabilities are not, strictly speaking, local observables, and they generally vary according to the mass representation chosen. The conditions under which the probabilities are left invariant by changes of mass representation are more restrictive than those required for covariance. For a complete discussion we refer to \cite{neut13}.    

\section{Applications}

We can now move on to apply the formulae \eqref{TransitionProbabilities2} to some spacetimes of interest. We skip the computation of the Minkowskian and quantum mechanical limits, which are trivial, and can be found in \cite{neut13}. For a specific metric, the computation essentially amounts to the determination of the Bogoliubov coefficients \eqref{BogoliubovCoefficients1}, which are then combined as in eq. \eqref{TransitionProbabilities2} to extract the transition probabilities. Then we are only limited by the knowledge of the solutions to the Dirac equations. There is a class of metrics (spatially flat Friedmann-Robertson-Walker (FRW) spacetimes) which are of interest in cosmology and for which there is often an exact analytical solution of the Dirac equation. The general form of the spatially flat FRW metric is, in a rectangular coordinate system,
\begin{equation}\label{FRWmetric}
 ds^2 = dt^2 - a^2 (t) \left( dx^2 + dy^2 + dz^2 \right) \ .
\end{equation}
The function $a(t)$ is known as the scale factor. According to the shape of $a(t)$, the metric of eq. \eqref{FRWmetric} describes various phases in the evolution of a isotropic, homogeneous and spatially flat universe. The spatial dependence of the solutions to the Dirac equations corresponding to the metric \eqref{FRWmetric} is trivial. Due to the spatial translation symmetry we can put
\begin{equation}\label{GenericSolution}
 \phi_{\pmb{k},s; j} (x) = u_{\pmb{k},s;j} (t) e^{i \pmb{k} \cdot \pmb{x}} \ ; \ \ \ \ \psi_{\pmb{k},s; j} (x) = v_{\pmb{k},s;j} (t) e^{i \pmb{k} \cdot \pmb{x}} \ .
\end{equation}
Here the generalized momentum index has been explicitly substituted with the $3$-vector $\pmb{k}$ and $\pmb{x}$ denotes the spatial $3$-vector $\pmb{x} \equiv (x,y,z)$. The $4$-component spinors $u_{\pmb{k},s;j} (t), v_{\pmb{k},s;j} (t)$ depend only on time. We can anticipate from eqs. \eqref{GenericSolution} that the Bogoliubov coefficients will have the form
\begin{equation}
 \Lambda_{\pmb{k},s; \pmb{q},r} (t) = \delta_{rs} \delta^3 (\pmb{k} - \pmb{q}) \Lambda_{k,s} (t) \ ; \ \ \ \  \Xi_{k,s; \pmb{q},r} (t) = \delta_{rs} \delta^3 (\pmb{k} + \pmb{q}) \Xi_{k,s} (t) \ .
\end{equation}
The natural choice for the foliation is given by the constant $t$ surfaces and we take the energy to be defined with respect to $\frac{\partial}{\partial t}$.
\begin{itemize}
\item $a(t) = e^{H t}$ . The exponential evolution is suited to describe inflation and epochs dominated by the cosmological constant. This form can also be considered as an approximate solution, for a suitable value of $H$, in time ranges where the Hubble rate $\frac{\dot{a}}{a}$ does not vary sensibly in time. The Hubble parameter $H$ has dimensions of an energy and its value depends on the epoch considered.  The solution of the Dirac equation for this scale factor can be found in \cite{Barut}. The corresponding Bogoliubov coefficients are{\small
\begin{eqnarray}\label{FRLWCOEFF1}
\nonumber  \Lambda_{k,s} (t) \! \! \! \! \! &=&  \! \! \! \! \! \frac{\pi k e^{-Ht}}{2 H \sqrt{\cos(\frac{i \pi m_2}{H}) \cos(\frac{i \pi m_1}{H})}}
  \left[ \!  J_{v_2}^*   \! \! \left( \! \! \frac{k}{H}e^{-Ht} \! \! \right) \! \!  J_{v_1} \! \!  \left( \! \! \frac{k}{H}e^{-Ht} \! \! \right) \! +\! J_{v_2 - 1}^*\! \! \left(\! \!  \frac{k}{H}e^{-Ht}\! \!  \right) \! \!  J_{v_1-1} \! \!  \left(\! \!  \frac{k}{H}e^{-Ht}\! \!  \right)\! \right]  \\ \nonumber
\Xi_{k,s} (t) \! \! \! \! \! &=&  \! \! \! \! \!  \frac{\pi k e^{-Ht}}{2 H \sqrt{\cos(\frac{i \pi m_2}{H}) \cos(\frac{i \pi m_1}{H})}}
 \left[ \!  J_{v_1}^*   \! \! \left( \! \! \frac{k}{H}e^{-Ht} \! \! \right) \! \!  J_{-v_2} \! \!  \left( \! \! \frac{k}{H}e^{-Ht} \! \! \right) \! - \! J_{v_1 - 1}^*\! \! \left(\! \!  \frac{k}{H}e^{-Ht}\! \!  \right) \! \!  J_{1-v_2} \! \!  \left(\! \!  \frac{k}{H}e^{-Ht}\! \!  \right)\! \right] \\ &&
\end{eqnarray}}
where $v_j = \frac
{1}{2}\left(1 + \frac{2 i m_j}{H}\right)$ and $J_{v} (x)$ denotes the Bessel function of the first kind of order $v$. Insertion in Eq. \eqref{TransitionProbabilities2} yields the result{\small
\begin{eqnarray}\label{ExponentialProbs}
 \nonumber && P^{e \rightarrow \mu}_{k,s}(t,t_0) =  2 \cos^{2}\theta \sin^{2}\theta\bigg\{1 -  \frac{\pi^2 k^2e^{-H(t+t_0)}}{4 H^2 \cos(\frac{i \pi m_2}{H}) \cos(\frac{i \pi m_1}{H})}  \\ \nonumber
&& \times \Re \bigg[ \left[  J_{v_2}    \left(  \eta_0  \right)   J_{v_1}^*  \left(  \eta_0 \right)  + J_{v_2 - 1} \left(  \eta_0  \right)   J_{v_1-1}^*    \left(  \eta_0 \right) \right]
\left[   J_{v_2}^*   \left( \eta  \right)   J_{v_1}   \left(  \eta  \right)  + J_{v_2 - 1}^* \left(  \eta  \right)   J_{v_1-1}   \left(  \eta  \right) \right]
\\  && + \left[   J_{v_1}    \left( \eta_0 \right)   J_{-v_2}^*  \left(  \eta_0 \right)  - J_{v_1 - 1} \left( \eta_0  \right)   J_{1-v_2}^*   \left(\eta_0  \right) \right]
  \left[   J_{v_1}^*   \left( \eta \right)   J_{-v_2}   \left( \eta \right)  - J_{v_1 - 1}^* \left(\eta \right)  J_{1-v_2}   \left(\eta  \right)\! \right] \bigg] \bigg\} \ ,
\end{eqnarray}}
with $\eta = \frac{k}{H}e^{-Ht}$ and $\eta_0 = \frac{k}{H}e^{-Ht_0}$. To understand the qualitative behaviour of Eq. \eqref{ExponentialProbs} we have plotted it for sample values of masses and momenta in the left panel of (Fig. 1). We notice that for small times the oscillations display an interference pattern similar to that of coupled harmonic oscillators. As time is increased, the transition probabilities gradually converge to a flat space--like oscillation and the interference pattern eventually disappears. In the right panel of (Fig. 1) we show a qualitative comparison with the Pontecorvo formulae, for different sample values of masses and momenta.

\begin{figure}\label{Figure1}
\centering
{\includegraphics[width=0.48 \linewidth]{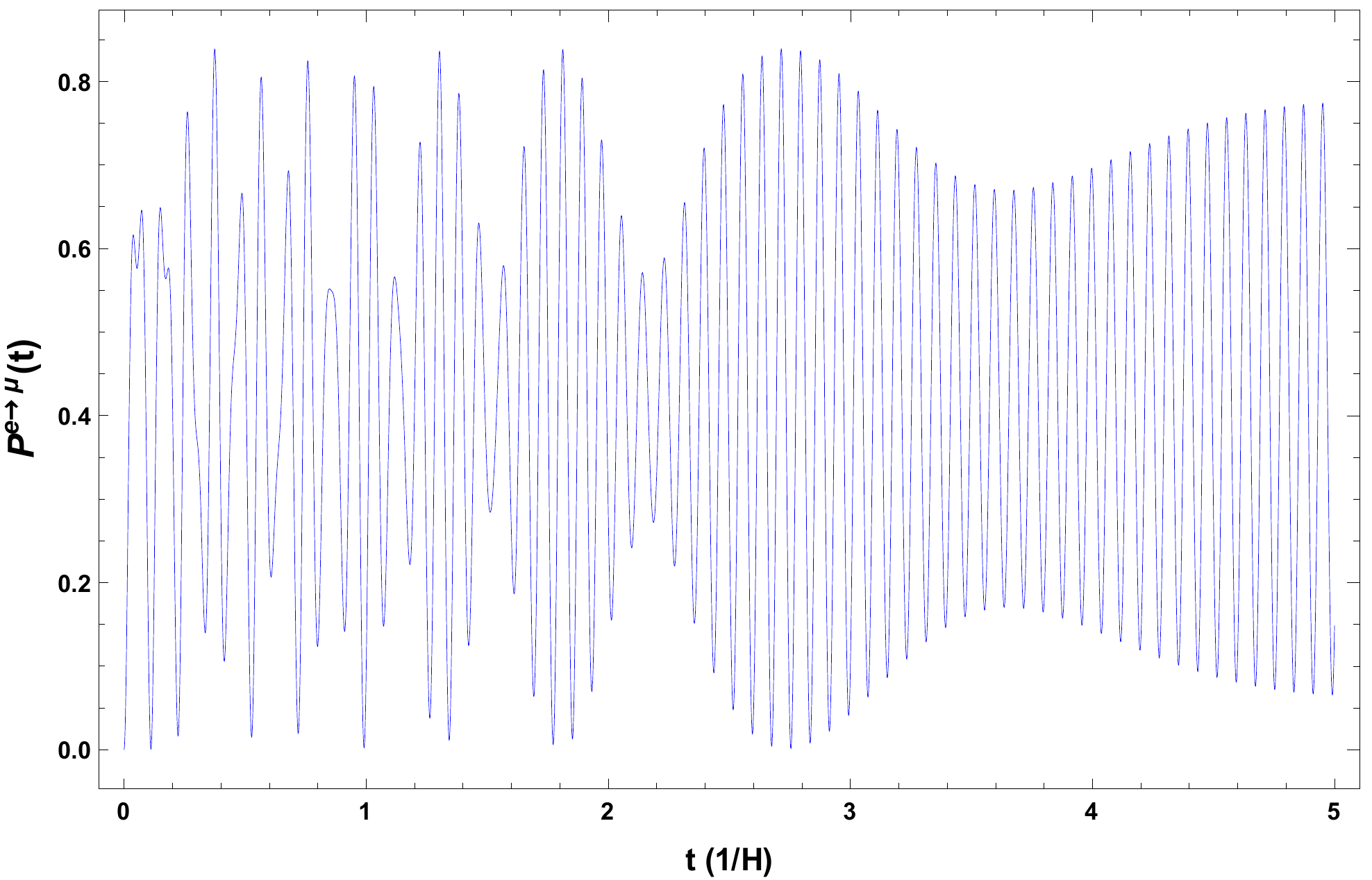}}
{\includegraphics[width=0.48 \linewidth]{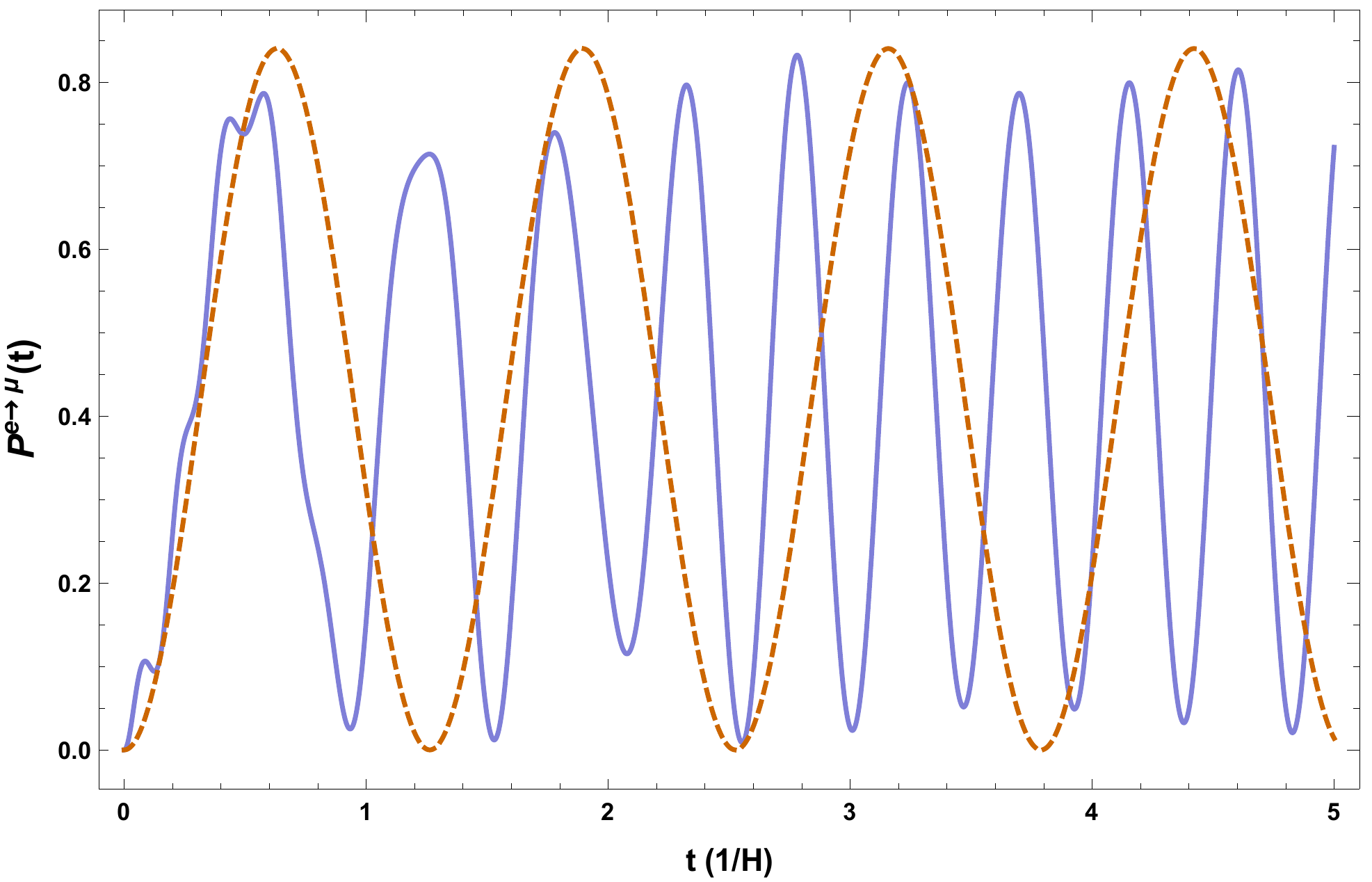}}
\caption{ Plots of the oscillation formulae from Eq. \eqref{ExponentialProbs} for different sample values of masses and momenta, chosen in order to highlight the qualitative behaviour of Eq. \eqref{ExponentialProbs}.  Masses and momenta are expressed in units of $H$, time is expressed in units of $H^{-1}$. (Left panel) Plot of the $\nu_e-\nu_{\mu}$ transition probability Eq. \eqref{ExponentialProbs} as a function of time. Here we have used the sample values $\sin^2(\theta) = 0.3, k = 30, m_1 =1 , m_2 = 80$ and $t_0 =0$. (Right panel) Plot of the oscillation formulae from Eq. \eqref{ExponentialProbs} (blue solid line) and from the Pontecorvo formulae (orange dashed line) for sample values of masses and momenta. Here we have used $\sin^2(\theta) = 0.3, k = 20, m_1 =1 , m_2 = 15$ and $t_0 = 0$.}
\end{figure}
\item $a(t) = a_0 t^{\frac
{1}{2}}$. This evolution of the scale factor corresponds to a universe dominated by radiation. The constant $a_0$ has dimensions of $[\mathrm{mass}]^{\frac
{1}{2}}$. The solution for the corresponding equation may be found in \cite{Barut} The mixing coefficients read

\begin{eqnarray}
  \nonumber \Lambda_{k,s} (t) &=&  \frac{1}{\sqrt[4]{4m_1m_2t^2}} e^{-\frac{\pi k^2 (m_1+m_2)}{4m_1m_2a_0^2}}
 \bigg\{ W^{*}_{\kappa_2,\frac{1}{4}} (-2im_2t) W_{\kappa_1,\frac{1}{4}} (-2im_1t)\\
 \nonumber  &+& \frac{4k^2}{m_1 m_2 a_0^2 t}  \bigg[\frac{1}{4}W^{*}_{\kappa_2, \frac{1}{4}} (- 2 im_2t) - \frac{1}{8}\left(1 + \frac{ik^2}{m_2 a_0^2} \right)W^{*}_{\kappa_2 - 1,\frac{1}{4}} (-2 i m_2 t) \bigg] \\
 &\times& \bigg[\frac{1}{4} W_{\kappa_1,\frac{1}{4}}(-2 im_1t) - \frac{1}{8} \left(1 - \frac{ik^2}{m_1 a_0^2} \right)W_{\kappa_1 -1,\frac{1}{4}}(-2 i m_1 t) \bigg] \bigg\}
\end{eqnarray}
\begin{eqnarray}
  \nonumber \Xi_{k,s} (t) &=&  (-1)^s  \frac{k}{\sqrt[4]{2m_1(2m_2)^3a_0^4t^2}} e^{-\frac{\pi k^2 (m_1+m_2)}{4m_1m_2a_0^2}}  \bigg\{ W^{*}_{\kappa_1,\frac{1}{4}} (-2im_1t) W_{-\kappa_2,\frac{1}{4}} (2im_2t)\\
\nonumber  &+& \frac{k^2}{m_1 m_2 a_0^2 t}
 \bigg[\frac{1}{4}W^{*}_{\kappa_1, \frac{1}{4}} (- 2 im_1t) - \frac{1}{8}\left(1 + \frac{ik^2}{m_1 a_0^2} \right)W^{*}_{\kappa_1 - 1,\frac{1}{4}} (-2 i m_1 t) \bigg]  \\  &\times& \bigg[ W_{-\kappa_2,\frac{1}{4}}(2 im_2t) + \frac{2im_2a_0^2}{k^2} W_{-\kappa_2 +1,\frac{1}{4}}(2 i m_2 t) \bigg] \bigg\}
\end{eqnarray}
where $W_{\kappa, \mu} (z)$ are the Whittaker functions \cite{Abramowitz} and \mbox{$\kappa_j = \frac{1}{4} \left(1 + \frac{2ik^2}{a_0^2 m_j} \right)$ for $j=1,2$}. The transition probabilities are 
{\small
\begin{eqnarray}
\nonumber && P^{e \rightarrow \mu}_{k,s}(t) =  2 \cos^{2}\theta \sin^{2}\theta\bigg\{1 -
 \Re \bigg[ \frac{1}{\sqrt[2]{4m_1m_2t_0 t}} e^{-\frac{\pi k^2 (m_1+m_2)}{2m_1m_2a_0^2}}
  \bigg\{ W_{\kappa_2,\frac{1}{4}} (-2im_2t_0) W^{*}_{\kappa_1,\frac{1}{4}} (-2im_1t_0)  \\
  \nonumber && + \frac{4k^2}{m_1 m_2 a_0^2 t_0}
 \left(\frac{1}{4}W_{\kappa_2, \frac{1}{4}} (- 2 im_2t_0) - \frac{1}{8}\left(1 - \frac{ik^2}{m_2 a_0^2} \right)W_{\kappa_2 - 1,\frac{1}{4}} (-2 i m_2 t_0) \right) \\ \nonumber && \times
 \left(\frac{1}{4} W^{*}_{\kappa_1,\frac{1}{4}}(-2 im_1t_0) - \frac{1}{8} \left(1 + \frac{ik^2}{m_1 a_0^2} \right)W^{*}_{\kappa_1 -1,\frac{1}{4}}(-2 i m_1 t_0) \right) \bigg\}  \bigg\{ W^{*}_{\kappa_2,\frac{1}{4}} (-2im_2t) W_{\kappa_1,\frac{1}{4}} (-2im_1t) \\
\nonumber && + \frac{4k^2}{m_1 m_2 a_0^2 t}
 \left(\frac{1}{4} W^{*}_{\kappa_2, \frac{1}{4}} (- 2 im_2t) - \frac{1}{8}\left(1 + \frac{ik^2}{m_2 a_0^2} \right)W^{*}_{\kappa_2 - 1,\frac{1}{4}} (-2 i m_2 t) \right) \\ \nonumber && \times
 \left(\frac{1}{4} W_{\kappa_1,\frac{1}{4}}(-2 im_1t) - \frac{1}{8} \left(1 - \frac{ik^2}{m_1 a_0^2} \right)W_{\kappa_1 -1,\frac{1}{4}}(-2 i m_1 t) \right) \bigg\} \\
\nonumber && +  \frac{k^2}{\sqrt[2]{2m_1(2m_2)^3a_0^4t_0 t}} e^{-\frac{\pi k^2 (m_1+m_2)}{2m_1m_2a_0^2}}  \bigg\{ W_{\kappa_1,\frac{1}{4}} (-2im_1t_0) W^{*}_{-\kappa_2,\frac{1}{4}} (2im_2t_0)
\end{eqnarray}
\begin{eqnarray}\label{WhittakerProbs}
\nonumber && + \frac{k^2}{m_1 m_2 a_0^2 t_0}
 \left(\frac{1}{4}W_{\kappa_1, \frac{1}{4}} (- 2 im_1t_0) - \frac{1}{8}\left(1 - \frac{ik^2}{m_1 a_0^2} \right)W_{\kappa_1 - 1,\frac{1}{4}} (-2 i m_1 t_0) \right) \\ \nonumber && \times
 \left( W^{*}_{-\kappa_2,\frac{1}{4}}(2 im_2t_0) - \frac{2im_2a_0^2}{k^2} W^{*}_{-\kappa_2 +1,\frac{1}{4}}(2 i m_2 t_0) \right) \bigg\}  \bigg\{ W^{*}_{\kappa_1,\frac{1}{4}} (-2im_1t) W_{-\kappa_2,\frac{1}{4}} (2im_2t) \\
 \nonumber
  && + \frac{k^2}{m_1 m_2 a_0^2 t}  \left(\frac{1}{4}W^{*}_{\kappa_1, \frac{1}{4}} (- 2 im_1t) - \frac{1}{8}\left(1 + \frac{ik^2}{m_1 a_0^2} \right)W^{*}_{\kappa_1 - 1,\frac{1}{4}} (-2 i m_1 t) \right) \\ && \times \left( W_{-\kappa_2,\frac{1}{4}}(2 im_2t) + \frac{2im_2a_0^2}{k^2} W_{-\kappa_2 +1,\frac{1}{4}}(2 i m_2 t) \right) \bigg\} \bigg] \bigg\} \ .
\end{eqnarray}
}
The expression for the transition probabilities, eq. \eqref{WhittakerProbs} is quite involved. In order to get some insight we plot the eq. \eqref{WhittakerProbs} for sample values of masses and momenta in (Fig. 2). 
\begin{figure}[h]
\includegraphics[width=18pc]{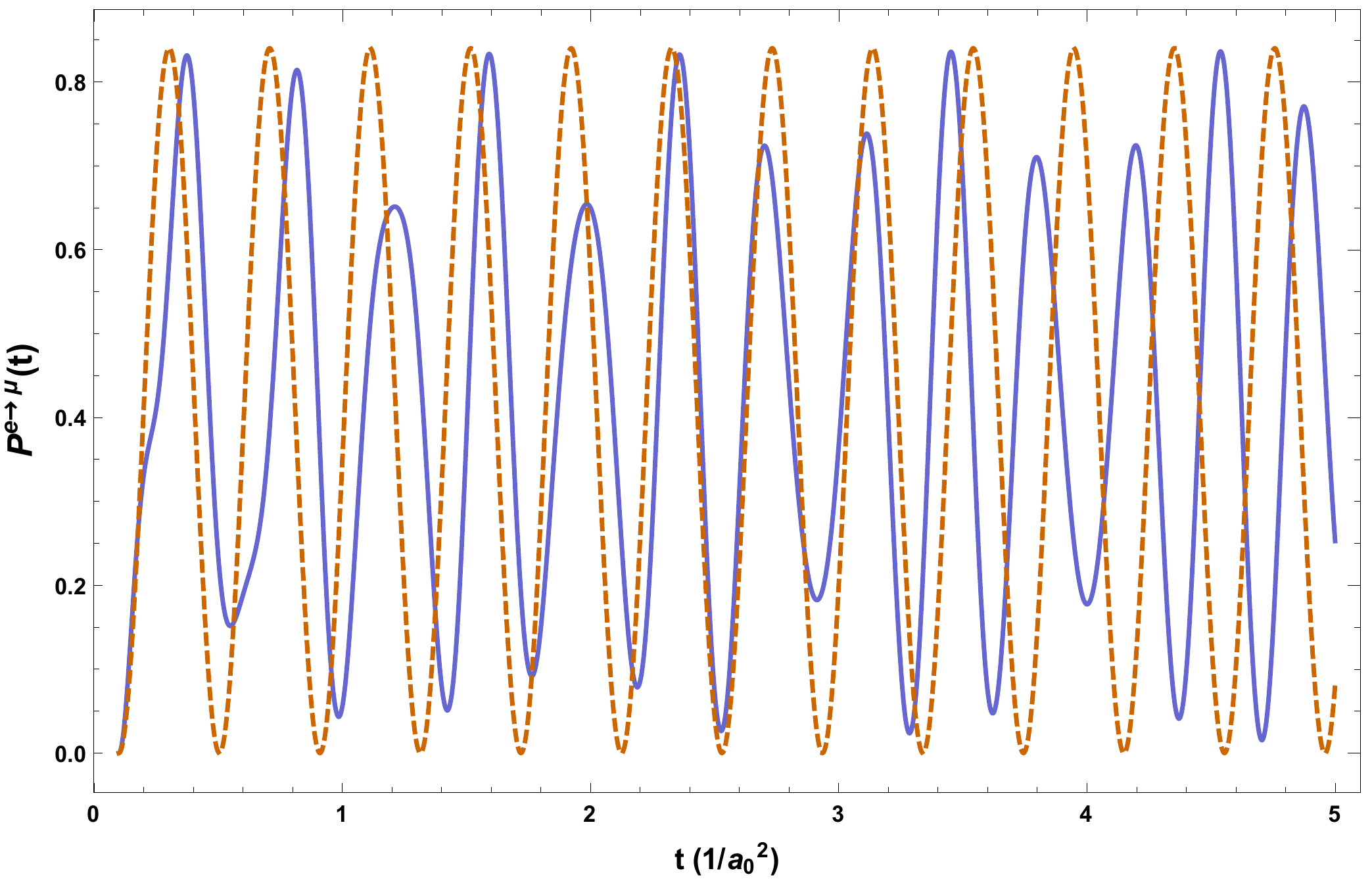}\hspace{2pc}%
\begin{minipage}[b]{18pc}\caption{\label{Figure2}Plots of the $\nu_e-\nu_{\mu}$ transition probability as a function of time as from Eq. \eqref{WhittakerProbs} (blue solid line) and from the Pontecorvo formulae (orange dashed line) for sample values of massed and momenta, chosen in order to highlight the qualitative behaviour of Eq. \eqref{WhittakerProbs}. Masses and momenta are expressed in units of $a_0^2$, time is expressed in units of $a_0^{-2}$. We have used the sample values $\sin^2(\theta) = 0.3, k = 5, m_1 =1 , m_2 = 20$ and $t_0 =0.1$.}
\end{minipage}
\end{figure}
Despite the analytical complexity, the probabilities of eq. \eqref{WhittakerProbs} share some common features with those of eq. \eqref{ExponentialProbs}. Both show amplitude and phase variations with respect to the (flat) Pontecorvo oscillation formulae. In particular the amplitude variations present in eqs. \eqref{WhittakerProbs} and \eqref{ExponentialProbs} are a distinctive feature of quantum field theory in curved spacetime. The latter cannot be obtained in the quantum mechanical limit, which in curved space modifies the Pontecorvo formulae only in the phase of the oscillations \cite{QMO3}.
\end{itemize}
As a further application we consider the (static) Schwarzschild metric
\begin{equation}
ds^2 = \left(1-\frac
{2 GM}{r}\right) dt^2 - \left(1-\frac
{2 GM}{r}\right)^{-1} dr^2 - r^2 d \Omega \ .
\end{equation}
Unfortunately the corresponding Dirac equation has no exact analytical solution. Nonetheless we can work out an approximate form of the transition probabilities for the propagation of neutrinos from the asymptotic past to the asymptotic future. We shall not give the details of the calculation here, and we refer to \cite{neut13} for a complete treatment. The trick is to consider the asymptotic solutions of the Dirac equation at the future and past infinities and the relations among them. One is then able to write down the Bogoliubov coefficients in terms of the asymptotic (flat) solutions. The final result is 
\begin{eqnarray}\label{HawkingProb}
\nonumber  &&  P^{e \rightarrow \mu}_{\omega} (m,n) \approx \\ \nonumber   && 2 \cos^2\theta \sin^2 \theta \
\bigg( 1   - \sqrt{\left[1 - F_H (\omega)\right] \left[1 - F_H (\omega')\right]}   \left[|U_{\omega;\omega'}|^2 \cos (\Delta^{-}_{\omega;m,n}) + |V_{\omega;\omega'}|^2 \cos(\Phi^{-}_{\omega;m,n})\right]  \\
\nonumber  &&+ \sqrt{F_H(\omega)\left[1 - F_H(\omega') \right]} |U_{\omega;\omega'}| |V_{\omega;\omega'}|   \left[ \cos (\Theta^{-}_{\omega;m,n}) - \cos(\Psi^{-}_{\omega;m,n})\right] \\
\nonumber &&+ \sqrt{F_H(\omega')\left[1 - F_H(\omega) \right]}|U_{\omega;\omega'}| |V_{\omega;\omega'}|  \left[ \cos(\Psi^{+}_{\omega;m,n})- \cos(\Theta^{+}_{\omega;m,n})\right]  \\
  &&- \sqrt{F_H(\omega)F_H(\omega')}  \left[|U_{\omega;\omega'}|^2\cos(\Delta^{+}_{\omega;m,n}) + |V_{\omega;\omega'}|^2 \cos(\Phi^{+}_{\omega;m,n})  \right] \bigg)  \ .
\end{eqnarray}
Here $m,n$ are integers labeling two families of surfaces approaching respectively the past and future timelike infinity (the limit $m,n \rightarrow \infty$ is understood). The phase factors depend on the details of the surfaces considered, as well as the energy. The quantities $U_{\omega \omega'}$ and $V_{\omega \omega'}$ denote the mixing Bogoliubov coefficients in flat space in terms of the energies $\omega, \omega' = \sqrt{\omega^2 + m_2^2 - m_1^2}$. Finally $F_H (\omega) = (1 + e^{\frac{\omega}{k_B T_H}})^{-1}$ is the Fermi-Dirac distribution at the Hawking temperature $T_H = \frac{1}{8 \pi G M}$. The most remarkable feature of eq. \eqref{HawkingProb} is the appearance of the Hawking temperature, which shows how Hawking radiation directly affects the propagation of mixed fermions on a black hole spacetime. The typical behavior of eq. \eqref{HawkingProb} is shown in (Fig. 3).

\begin{figure}[h]
  \centering
  \includegraphics[width=0.5\linewidth]{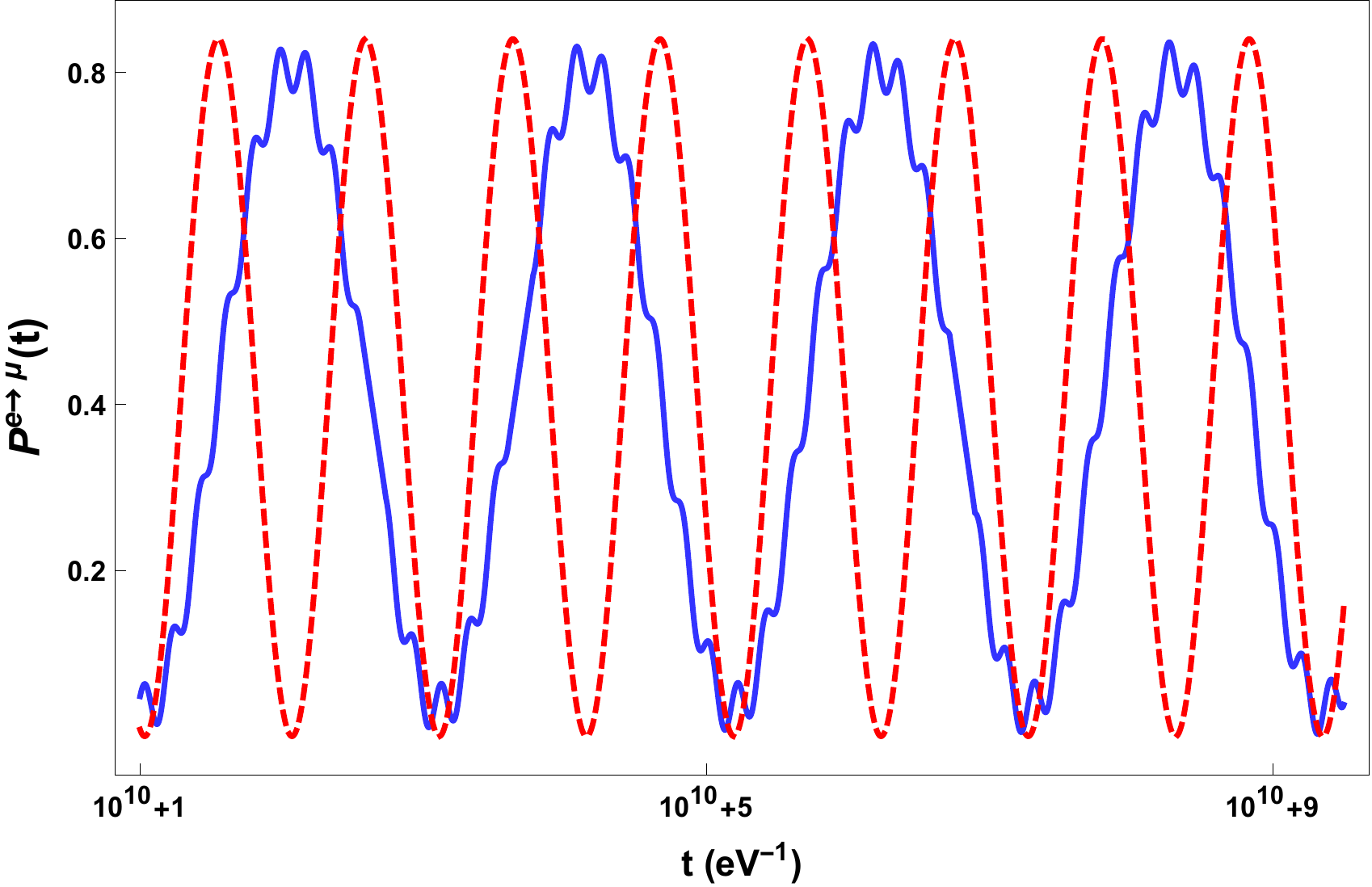}
  \caption{ Plot of the $\nu_{e} - \nu_{\mu}$ transition probability from Eq. \eqref{HawkingProb} (blue solid line) and from the Pontecorvo oscillation formulae (red dashed line) for late times and sample values of masses, momenta and Hawking Temperature. The phases in \eqref{HawkingProb} have been chosen so as to match the flat space phases for simplicity, $\Delta^{\pm}_{\omega;m,n} \rightarrow \frac{\omega_2 - \omega_1}{2} (t \pm t_0)$, $\Phi^{\pm}_{\omega;m,n}\rightarrow \frac{\omega_2 + \omega_1}{2} (t \pm t_0)$, $\Psi^{\pm}_{\omega;m,n} \rightarrow \frac{\omega_2}{2} (t \pm t_0) + \frac{\omega_1}{2}(t \mp t_0)$, $\Theta^{\pm}_{\omega;m,n} \rightarrow  \frac{\omega_2}{2} (t \pm t_0) - \frac{\omega_1}{2}(t \mp t_0)$, where $t$ and $t_0$ denote respectively the future and past hypersurfaces. We have used the sample values $\sin^2(\theta) = 0.3$, $k= 30 \ \mathrm{eV}$, $m_1 = 1  \ \mathrm{eV}$, $m_2 = 20  \ \mathrm{eV}$, $t_0 = 0$, $k_B T_H = 10^{-10} \ \mathrm{eV}$ and $t$ in the range $[10^{10}+1,10^{10}+9.5] \ \mathrm{eV}^{-1}$. }
\end{figure}

\section{Conclusions}
We have constructed the quantum field theory of (two flavor) fermion mixing in curved spacetime. We have derived general oscillation formulae and applied them to several spacetimes of interest. The theory has a considerably richer structure when compared to its flat spacetime and its quantum mechanical counterparts, and joins the peculiarities of field mixing with the inherent ambiguity of field quantization in curved spacetime. Such a complexity is mirrored in the oscillation formulae, which involve a non-trivial evolution in both phase and amplitude. A similar analysis can also be set up for boson mixing \cite{BM1,BM2} The formalism developed here is suited for the analysis of neutrino oscillations in situations where gravity plays an important role. That includes extreme astrophysical environments, such as the primordial universe and black holes. Moreover the theory allows one to study the flavor vacuum in an arbitrary spacetime. Due to its condensate structure, the flavor vacuum yields a non-vanishing contribution to the energy momentum tensor of matter, inducing a new source term on the right hand side of the Einstein field equations. Recent developments \cite{DM4} have shown that in cosmological spacetimes the flavor vacuum behaves as a perfect fluid with the dust (or cold dark matter) equation of state $w=0$. This gives rise to the intriguing possibility that a pure field theoretical effect may contribute to dark matter. Future studies shall explore the link between field mixing and dark matter in a wider class of metrics.    

\section*{Acknowledgements}
Partial financial support from MUR and INFN is acknowledged. A.C. and G.L. also acknowledge the COST Action CA1511
Cosmology and Astrophysics Network for Theoretical Advances and Training Actions (CANTATA).

\end{document}